\documentclass[10pt]{iopart}
\usepackage{iopams}  
\usepackage{amssymb}
\usepackage{setstack}
\usepackage{graphicx}
\usepackage{dcolumn}
\usepackage{bm}
\usepackage{hyperref}
\usepackage{braket}
\usepackage{comment}

\usepackage{comment}
\usepackage{bbm}
\usepackage{cite}

\begin{document}

\title[Fubini-Study metric and topological properties]{Fubini-Study metric and topological properties of flat band electronic states: the case of an atomic chain  with $s-p$ orbitals}

\author{Abdiel Espinosa-Champo and Gerardo G. Naumis.}

\address{Depto. de Sistemas Complejos, Instituto de F\'isica, Universidad Nacional Aut\'onoma de M\'exico (UNAM). Apdo. Postal 20-364, 01000, CDMX, Mexico.}
\ead{naumis@fisica.unam.mx}
\vspace{10pt}
\begin{indented}
\item[]\today
\end{indented}

\begin{abstract}
The topological properties of the flat band states of a one-electron Hamiltonian that describes a chain of atoms with $s-p$ orbitals are explored. This model is mapped onto a Kitaev-Creutz type model, providing a useful framework to understand the topology through a nontrivial winding number and the geometry introduced by the \textit{Fubini-Study (FS)}  metric.  This metric allows us to distinguish between pure states of systems with the same topology and thus provides a suitable tool for obtaining the fingerprint of flat bands. Moreover, it provides an appealing geometrical picture for describing flat bands as it can be
associated with a local conformal transformation over circles in a complex plane. In addition, the presented model allows us to relate the topology with the formation of Compact Localized States (CLS) and pseudo-Bogoliubov modes.  Also, the properties of the squared Hamiltonian are investigated in order to provide a better understanding of the localization properties and the spectrum. The presented model is equivalent to two coupled SSH chains under a change of basis. 
\end{abstract}

\maketitle

\section{Introduction\label{Sec:Introduction}}
A flat band refers to a band with constant energy unaffected by the crystal momentum. This property suppresses wave transport and makes it highly sensitive to perturbations \cite{Leykam_Andreanov_Flach_2018}. This has led to the exploration of partially flat bands, which have vanishing dispersion along specific directions or near particular points in the Brillouin zone \cite{Bergholtz_Liu_2013, Nguyen2018, DENG2003412}.

Due to their unique characteristics, flat band systems have been a subject of great interest in several research fields \cite{QiuWenXuan, Drost_Ojanen_Harju_Liljeroth_2017, Abilio1999, HidekiOzawa2015, Nakata2012, He_Mao_Cai_Zhang_Li_Yuan_Zhu_Wang_2021, Leonardo2021, Leonardo2022}, such as the generation of electronic correlations in condensed matter \cite{Mielke_Tasaki_1993, Tasaki_1998, Cao2018, AokiHideo2020, WuCongjun2007, Jaworowski_2018}; among such phenomena include ferromagnetism \cite{Mielke_Tasaki_1993, Tasaki_1998}, superconductivity \cite{Cao2018, AokiHideo2020}, and Wigner crystal formation \cite{WuCongjun2007, Jaworowski_2018}, and in photonics leading to slow-light realizations \cite{Settle:07, Krauss2007} and coherent propagation free of quantum dispersion \cite{Valagiannopoulos2019Feb, Valagiannopoulos2019Mar}. Thus, the study of flat band systems is crucial because it provides insight into the collective phenomena that govern the behavior of complex materials\cite{A_Mielke_1991, Tasaki1992, Mielke_Tasaki_1993, AokiHideo1993, AokiHideo1996}. These materials are of great interest because they have the potential to revolutionize many areas, such as electronics and possible applications to quantum computing \cite{AlexanderKerlsky2021}. 
 
Historically, developing flat band models has been a long and arduous process. It started with Sutherland's discovery of a flat band in the dice lattice \cite{Sutherland1986}. It continued with Lieb's work on the Hubbard model, demonstrating that certain bipartite lattices with chiral flat bands exhibit ferromagnetism \cite{Lieb1989}. However, in recent years,  there has been a growing interest in the development of new topological flat band models \cite{Bergholtz_Liu_2013} that will allow for a better understanding of the properties of these materials and their potential applications and which can support quantum Hall-like states, including integer quantum Hall (QH) effect \cite{Liu_2014, Thouless1982}, fractional quantum Hall (FQH) effect \cite{Bernevig2011, PARAMESWARAN2013816, ZhiLi2018}, and the existence of electronic fractional Chern states \cite{Laughlin1983, Bernevig2011, Leonardo2021, Leonardo2022}. 

Lastly, one of the challenges in studying flat band systems is distinguishing between pure states of systems with the same topology. To overcome this challenge, in previous work, the FS metric has been introduced as a tool, mainly to differentiate quantum states in flat bands \cite{LEDWITH2021168646, TomokiOzawa2021Sep, TomokiOzawa2021Jul, TomokiOzawa2021Jul13}. This metric enables the reliable identification of flatness regions in topological systems.

This paper presents a non-superconducting one-dimensional tight-binding model that can be mapped to a Kitaev chain Hamiltonian, preserving its topological properties with a nontrivial winding number. The \textit{FS} metric of the model allows for the construction of a mapping $f$ that can accurately distinguish between topological and nontopological systems, as well as between topological systems with and without flat bands. This model is inspired by recent experimental evidence of one-dimensional flat bands along established directions in two-dimensional van der Waals structures \cite{Yafei2023}, as well as research that suggests that chains of elements such as boron \cite{ChaiJengDa2018}, gallium \cite{Vidya2018} and tellurium \cite{Mohd2020, Harrison_1989} could be used to realize the proposed model experimentally.

This paper is organized as follows. Section \ref{Sec:Model} introduces the atomic chain model and the effective Hamiltonian. Section \ref{Sec: General properties} discusses the formation of pseudo-Bogoliubov modes and a regime with flat bands where compact localized states (CLS) exist. These bands are characterized by relations similar to those found in Landau levels. Additionally, we demonstrate a nontrivial topology phase transition and how the geometry described by the FS metric enables the mapping $f$ to be constructed to differentiate between flat band regimes and other topological systems. Finally, Section \ref{Sec: Conclusions} summarizes our findings.

\section{Model and Methods \label{Sec:Model}}
\begin{figure}
    \centering
    \large{a)}\includegraphics[scale=0.5]{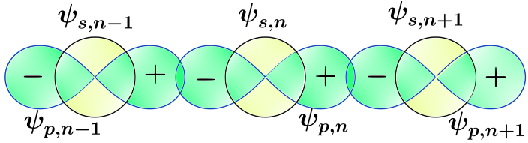}\\
    \large{b)} \includegraphics[scale=0.7]{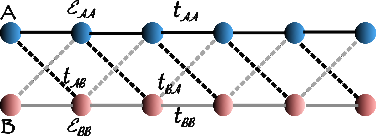}
    \caption{a) At each site of the atomic chain, there are two orbitals $s$ and $p_{x}$, denoted as $\psi_{s,n}$ and $\psi_{p,n}$. The hopping integrals of the nearest-neighbor $t_{pp}$ and $t_{ss}$ have the opposite sign, while the nearest-neighbor $t_{sp}$ couplings alternate in sign. Due to inversion symmetry, the on-site $t_{sp}$ coupling will be zero, just like in an isolated atom. The tight-binding Hamiltonian is given by Eq. \ref{eq:Model.1}. b) This atomic chain can be geometrically represented as an unbalanced Creutz ladder \cite{junemann_exploring_2017, He_Mao_Cai_Zhang_Li_Yuan_Zhu_Wang_2021, kuno_flat-band_2020, kang_creutz_2020, mukherjee_tailoring_2022}, where the blue and red circles represent the sites $A$ and $B$, respectively. The orbitals $s$ and $p$ correspond to the sites $A$ and $B$, and the hoppings follow the rules $t_{ss} \to t_{AA}$ (black lines), $t_{pp} \to t_{BB}$ (gray lines), $t_{sp (ps)} \to t_{AB (BA)}$ (dashed black (gray) lines) and $\varepsilon_{s(p)} \to \varepsilon_{A(B)}$ (see Eq. \ref{eq:Model.2}). }
    \label{fig:sporbitals}
\end{figure}

As explained in the introduction, our main motivation here is to find a model with realistic features such that it contains a flat band susceptible of being treated analytically in a simple way. Flat bands are associated with zero group velocity and this requires destructive wave interference. Clearly, a model based only on pure $s$ orbitals is  not able to produce such effect. The change of sign induced by $p$ orbitals when rotated by an angle of $\pi$ induces such possibility as positive and negative interactions of the same magnitude appear. Therefore, the most simple model is to have a one-dimensional system with   $s-p$ orbitals (see Fig. \ref{fig:sporbitals} a) ). Notice that here we do not introduce $p_y$ and $p_z$ orbitals due to several reasons. The first  is that we want to keep the model simple to shine light in the Fubini-Study metric topological study. But there are other physical reasons to proceed in such a way. One is that such simple system can be implemented using quantum analogous systems like in ultracold atom lattices\cite{junemann_exploring_2017} or simulate topological zero modes (flat bands) on a qubit superconducting processor \cite{qubit}. Having only one type of $p$ orbitals simplify considerably the complexity of the device. The second reason is that chalcogenide elements as $Se$ or $Te$ form chains \cite{Harrison_1989}. The bonds are directed along the chain direction and thus are well described with only one type of $p$ orbitals.

Under such considerations, our investigation is based on a tight-binding model with only first-neighbors hopping, which yields a Hamiltonian that can be expressed as

\begin{eqnarray} \label{eq:Model.1}
\mathcal{\hat{H}}= \sum_{\boldsymbol{r}_{n}}\left( \varepsilon_{s}(\boldsymbol{r}_{n}) c_{s, \boldsymbol{r}_{n}}^{\dag}c_{s, \boldsymbol{r}_{n}}+\varepsilon_{p}(\boldsymbol{r}_{n}) c_{p, \boldsymbol{r}_{n}}^{\dag}c_{p, \boldsymbol{r}_{n}} \right) \nonumber \\
 +\sum_{\boldsymbol{r}_{n}} \left[ t_{ss}(\boldsymbol{r}_{n}) c_{s, \boldsymbol{r}_{n+1}}^{\dag}c_{s, \boldsymbol{r}_{n}}+ t_{pp}(\boldsymbol{r}_{n}) c_{p, \boldsymbol{r}_{n+1}}^{\dag} c_{p, \boldsymbol{r}_{n}}\right. \nonumber\\
 \left. + t_{sp}(\boldsymbol{r}_{n}) c_{p, \boldsymbol{r}_{n+1}}^{\dag} c_{s, \boldsymbol{r}_{n}}+t_{ps}(\boldsymbol{r}_{n}) c_{s, \boldsymbol{r}_{n+1}}^{\dag} c_{p, \boldsymbol{r}_{n}} \right]+h.c.  
\end{eqnarray}
The indices $s$ and $p$ refer to the respective orbitals, and $\boldsymbol{r_{n}}$ is the position of the $n-th$ atom. The fermionic annihilation (creation) operator for orbital $s$ and $p$ is denoted by $c_{(s,p), \boldsymbol{r}_{n}} (c_{(s,p), \boldsymbol{r}_{n}}^{\dag})$, while $\varepsilon_{(s,p)}(\boldsymbol{r}_{n})$ and $t_{ss} (\boldsymbol{r}_{n}), t_{pp} (\boldsymbol{r}_{n}),t_{sp} (\boldsymbol{r}_{n}), t_{ps} (\boldsymbol{r}_{n})$ are the energy on-site and the hopping parameter to the first right neighbor, respectively. Assuming that $\varepsilon_{\alpha} (\boldsymbol{r}_{n})$ and $t_{\alpha, \beta}(\boldsymbol{r}_{n})$ are independent of the atomic position, the tight-binding parameters $t_{ss},t_{pp},t_{sp},t_{ps}$ can be obtained.

We can think of the chain as a ladder, with each $s$ and $p$ orbital mapped to different sites. As illustrated in Fig. \ref{fig:sporbitals} b), the ladder consists of two types of sites: type A, which is derived from the $s$ orbitals and type B, which is derived from the $p$ orbitals. This ladder is equivalent to a Creutz model, and its Hamiltonian is given by.

\begin{eqnarray} \label{eq:Model.2}
 \mathcal{H}= \sum_{n} \left( \varepsilon_{A} a_{n}^{\dag}a_{n}+ \varepsilon_{B} b_{n}^{\dag} b_{n} \right)\nonumber \\
   + \sum_{n} \left( t_{AA} a_{n+1}^{\dag} a_{n}+ t_{BB} b_{n+1}^{\dag} b_{n}+ t_{AB} b_{n+1}^{\dag} a_{n}+ t_{BA} a_{n+1}^{\dag} b_{n}\right)+ h.c.
\end{eqnarray}
The annihilation (creation) operators for sites $A$ and $B$ in the cell $n$ are indicated by $a_{n} (a_{n}^{\dag})$ and $b_{n} (b_{n}^{\dag})$, respectively. It has been demonstrated that due to the symmetry of the $p$ orbitals, $t_{ps}=-t_{sp}$ and $t_{pp}=-t_{ss}$ \cite{girvin_yang_2019}; consequently $t_{AA}= -t_{BB}$ and $t_{AB}=-t_{BA}$. In addition to this, we assume that $\varepsilon_{AA}= - \varepsilon_{BB}$, then the Hamiltonian is given by

\begin{eqnarray} \label{eq:Model.3}
    \mathcal{H}=  \varepsilon \sum_{n} \left( a_{n}^{\dag}a_{n}-  b_{n}^{\dag} b_{n} \right)  \nonumber \\
     +  \sum_{n} t_{AA}\left(  a_{n+1}^{\dag} a_{n}-  b_{n+1}^{\dag} b_{n} \right)+ t_{AB}\left(  b_{n+1}^{\dag} a_{n}- a_{n+1}^{\dag} b_{n}\right) + h.c. 
\end{eqnarray}
We then perform a lattice Fourier transformation using the operators
 \begin{equation}
     a_{k}= \frac{1}{\sqrt{N}} \sum_{n} a_{n} e^{i k x_{n}} \mbox{ and } b_{k}= \frac{1}{\sqrt{N}} \sum_{n} b_{n} e^{i k x_{n}},
 \end{equation}
where $x_{n}= n l$ and $l$ is the lattice constant. Finally, this allows us to rewrite the Hamiltonian Eq. \ref{eq:Model.3} in the standard Bogoliubov- de Gennes form.
\begin{eqnarray} \label{eq:Model.5}
 \mathcal{H}= t_{AA} \sum_{k} \boldsymbol{\Psi}_{k}^{\dag} \mathcal{H}(k) \boldsymbol{\Psi}_{k}, \,\,  \boldsymbol{\Psi}_{k}= (a_{k}, b_{k})^{T}.
\end{eqnarray}
where,
\begin{equation}
\mathcal{H}(k)= \boldsymbol{n(k) \cdot \sigma}=n_{x}(k) \sigma_{x}+n_{y}(k) \sigma_{y}+ n_{z}(k) \sigma_{z}
\label{eq:mostsimpleandfundamental}
\end{equation}
and the coefficients that accompany the Pauli matrices $\sigma_x,\sigma_y,\sigma_z$ are $n_{x}(k)=0, n_{y}(k)=  2 \lambda \sin(k l), n_{z}(k)= (\overline{\varepsilon}+2 \cos(k l))$. Here, $\overline{\varepsilon}$ and $\lambda$ are dimensionless parameters that capture the information of the parameters $\varepsilon$ and $t_{AB}$ of the Hamiltonian \ref{eq:Model.3}, respectively, and are defined as $\overline{\varepsilon}\equiv \varepsilon/t_{AA}$ and $\lambda \equiv t_{AB}/t_{AA}$.

 \section{Results\label{Sec: General properties}}

In this section, we explore the properties of the model proposed in Eq. \ref{eq:Model.5}. We examine the emergence of Bogoulibov and Majorana pseudo modes, their equivalence with the Kitaev Hamiltonian, their topological properties demonstrated by a nontrivial winding number, and the presence of a regime with topological flat bands and similarities to Landau levels. Additionally, we show that this system is equivalent to two coupled SSH chains and two decoupled chains with the next-nearest neighbor hopping, and we introduce a conformal transformation that allows us to identify topological and nontopological regimes and between flat and dispersive bands.

\subsection{Pseudo-Bogoliubov and Majorana modes}

We can define two pseudo-Bogoliubov modes, $\gamma_{k}$ and $\rho_{k}$, using the Hamiltonian in Equation \ref{eq:Model.5}. These modes are a combination of a fermion at sites $A$ and $B$ and are analogous to the Bogoliubov quasiparticles. They are expressed as 
\begin{eqnarray}
    \gamma_{k} \equiv u_{k} a_{k}+ v_{k} b_{k}, \,\, \varrho_{k} \equiv -v_{k}^{*} a_{k} +u_{k}^{*} b_{k},
\end{eqnarray}
where $u_{k}$ and $v_{k}$ are the coefficients that define the Bogoliubov modes. These pseudo-Bogoulibov modes hybridize orbitals $s$ and $p$, and satisfy the fermionic creation and annihilation anti-commutation relations (see \ref{Sec:Appendix A}).
\begin{eqnarray}\label{eq: Results.A.2}
\{ \varrho_{k}, \varrho_{k'}^{\dag}\}= \delta_{kk'};\,\, \{\varrho_{k}^{\dag}, \varrho_{k'}^{\dag}\}= \{ \varrho_{k}, \varrho_{k'} \}=0 ;\nonumber \\
 \{ \gamma_{k}, \gamma_{k'}^{\dag}\}= \delta_{kk'};\,\, \{\gamma_{k}^{\dag}, \gamma_{k'}^{\dag}\}= \{ \gamma_{k}, \gamma_{k'} \}=0
 \end{eqnarray}
where $u_{k}$ and $v_{k}$ must meet the criteria of $ u_{k} ^{2}+ v_{k} ^{2}=1$, $u_{-k}=u_{k}$ and $v_{-k}=- v_{k}$. A suitable selection of $u_{k}$ and $v_{k}$ will satisfy these conditions.
\begin{eqnarray} \label{eq:Results.A.3}
    u_{k}= \cos\left(\frac{\omega_{k}}{2} \right), v_{k}=-i \sin\left( \frac{\omega_{k}}{2} \right),
 \end{eqnarray}
 where we defined,
 \begin{equation}
     \omega_{k}= \mbox{Arg}\{2 i \lambda \sin(k l)+ (\overline{\varepsilon}+2 \cos(k l))\}
 \end{equation}
Here, $\mbox{Arg}(z)$ refers to the principal value of $z \in \mathbb{C}$. When substituting $a_{k}, b_{k}$ into Eq. \ref{eq:Model.5}, we diagonalize the Hamiltonian to obtain
\begin{eqnarray} \label{eq: Results.A.4}
 \fl    \mathcal{H}= t_{AA} \sum_{k} \epsilon(k) \left( \gamma_{k}^{\dag} \gamma_{k}- \varrho_{k}^{\dag} \varrho_{k} \right) \mbox{ with } \epsilon(k)= \sqrt{(\overline{\varepsilon}+2 \cos(kl))^{2}+(2 \lambda \sin(kl))^{2}}.
 \end{eqnarray}
Generally, these pseudo-Bogoliubov modes are associated with squeezed coherent states \cite{Thirulogasanthar2018} and, in a similar form, have recently been observed in twisted bilayer graphene (TBLG) at magic angles \cite{LeonardoLabastida2022}. 

\begin{figure}[t]
    \centering
    \includegraphics[scale=0.6]{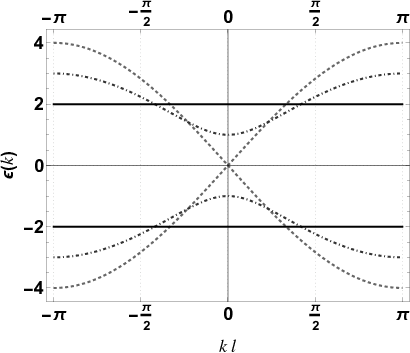}
    \caption{The band structure of the Kitaev-Creutz ladder model as a function of $kl$, and for the set of values $\lambda=1$, $l=1$, and $\overline{\varepsilon}=-2$ (dotted line), $\overline{\varepsilon}=-1$ (dot-dashed line), and $\overline{\varepsilon}=0$ (solid line), respectively (see Eq. \ref{eq: Results.A.4}).}
    \label{fig:energy-spectra}
\end{figure}

In the long-wavelength limit, we can expand the Hamiltonian \ref{eq:Model.5}  at $k=0$ to obtain 
\begin{eqnarray} \label{eq: Results.A.7}
 \fl   \mathcal{H}= t_{AA} \sum_{k} \boldsymbol{\Psi}_{k}^{\dag}H_{D}(k) \boldsymbol{\Psi}_{k} \mbox{ with } H_{D}(k)= m \sigma_{z}+ 2 \lambda k l \sigma_{y} \mbox{ and } m= (\overline{\varepsilon}-\overline{\varepsilon}_{c}) .
\end{eqnarray}
where $\overline{\varepsilon}_{c}=-2$ and the energy dispersion is given by $\epsilon(k)= \pm \sqrt{(\overline{\varepsilon}-\overline{\varepsilon}_c)^{2}+ 4 \lambda^{2} k^{2} l^{2}}$.
As shown in Figure \ref{fig:energy-spectra}, there is an energy gap of size $\Delta=2(\overline{\varepsilon}-\overline{\varepsilon}_c)$ that vanishes when the critical value $\overline{\varepsilon}_c$ is reached, that is, when $m$ is close to zero. At this point, the energy dispersion follows the relation $\epsilon(k)= \pm 2 \lambda l k $, implying that the pseudo-Bogoliubov modes can be interpreted as pseudo-Majorana modes, and they can move along the chain with a velocity of $ v = 2 \lambda l$. As the mass approaches zero, the energy of these eigenstates is equal.

\subsection{Topological properties I: Nontrivial winding number}
\begin{figure}[t]
    \centering
    \large{a)}\includegraphics[scale=0.5]{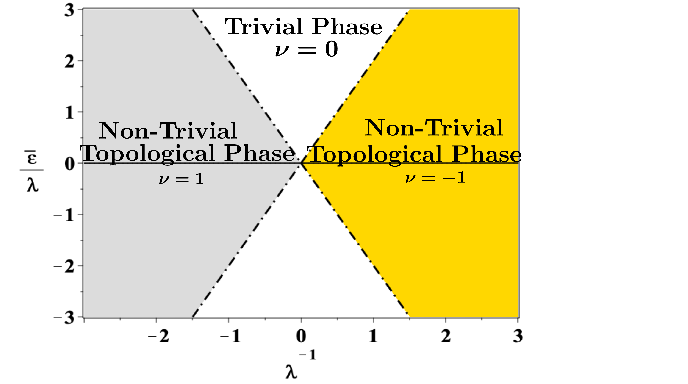}   \\
    \large{b)} \includegraphics[scale=0.4]{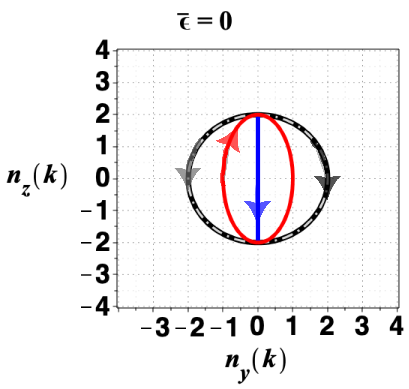}
    \large{c)} \includegraphics[scale=0.4]{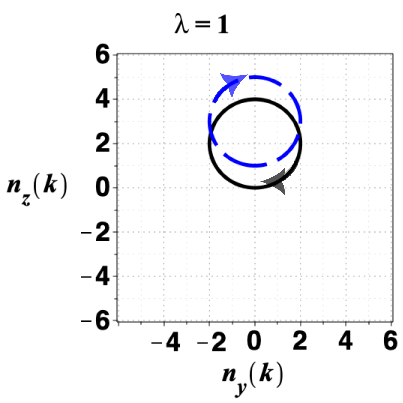}
   \caption{The topological phase diagram of the Kitaev-Creutz ladder system with Hamiltonian \ref{eq:Model.5} is shown in Fig. a) using Eq. \ref{eq:Results.B.1}. The dashed black lines indicate the boundaries between the topological phases, which are determined by the condition $\overline{\varepsilon}= \pm 2$. The curves in panels b) and c) correspond to the parameterization $(n_{y}(k), n_{z}(k))$ (cf. Eq. \ref{eq:Model.5}), where $k \in \mbox{BZ}$. In Fig. b), the curves for $\lambda=-1$ (gray circle), $\lambda=0$ (blue line), $\lambda=1/2$ (red ellipse), and $\lambda=1$ (black circle) are plotted with $\overline{\varepsilon}=0$. In Fig. c), the curves for $\overline{\varepsilon}=2$ (black circle) and $\overline{\varepsilon}=3$ (blue dashed circle) are plotted with $\lambda=1$. It is noteworthy that for the blue line in Fig. b) and the black circle in Fig. c), the winding number is not well defined since the curve passes through the origin. }
    \label{fig:phase-diagram}
\end{figure}
The Hamiltonian \ref{eq:Model.5} can be exactly mapped to a Kitaev Hamiltonian, extensively studied by Leumer et al. \cite{Leumer_2020}. Hence, our model is referred to as the Kitaev-Creutz model and the correspondence is as follows
\begin{eqnarray} \label{eq:correpondence-to-Kitaev}
     \overline{\varepsilon} \to -\mu/t, \,\,\, \lambda \to - \Delta/ t,
 \end{eqnarray}
however, the physics interpretation is not the same. 
It has been demonstrated in \cite{Leumer_2020} that the Hamiltonian \ref{eq:Model.5} is invariant under time-reversal symmetry for spinless fermions $\mathcal{T}= \mathbbm{1} \mathcal{K}$, with $\mathcal{K}$ being the complex conjugation and the chiral symmetry $\mathcal{TP}= \mathcal{C}= \sigma_{x}$. Additionally, it anti-commutes with the particle-hole operator $\mathcal{P}= \sigma_{x} \mathcal{K}$.  
 Therefore, the particle-hole symmetry establishes that the band structure is symmetric with respect to the zero energy. Note that the Kitaev Hamiltonian belongs to the BDI class \cite{Altland1997}, where all square symmetries operators are the identity.

The chiral symmetry allows us to define the winding number as the topological invariant \cite{Kempkes2019}, where the winding number is defined as \cite{Kempkes2019,Chiu2016} 
\begin{eqnarray} \label{eq:Results.B.1}
    \nu= \frac{1}{2 \pi} \int_{- \pi/l}^{\pi/l} dk \partial_{k} \omega_{k},
\end{eqnarray}
here $\partial_{k} \omega_{k}$ is the winding number density \cite{Kempkes2019,Leumer_2020}. The phase diagram in Fig. \ref{fig:phase-diagram} is similar to that of a Kitaev chain with the appropriate parameters \cite{Leumer_2020}. The dashed black lines in Fig. \ref{fig:phase-diagram} a) indicate the boundaries between the topological phases with $\overline{\varepsilon}= \pm 2$, which meet the condition $\epsilon(k)=0$ in $kl=0, \pm \pi d$ and $\lambda \neq 0$. Figures \ref{fig:phase-diagram} b)-c) show the curves of parameterization $(n_{y}(k), n_{z}(k))$ (cf. Eq. \ref{eq:Model.5}) along the Brillouin zone (BZ) with different values of $\overline{\varepsilon}$ and $\lambda$, with the winding number being the topological invariant.


\subsection{Flat bands, Compact Localized States (CLS) and Analogous Landau Levels Relations \label{Sec: Analogous_Landa_Levels}}
When a flat band is present, the group velocity of the charge carriers is zero for all momenta in the Brillouin zone, indicating that the charge carriers are localized. This localization is caused by the presence of a particular localized eigenstate, known as the compact localized state (CLS). This state has a finite amplitude within a finite region in real space and is zero outside. It should be noted that CLS is not unique and can be of multiple types, depending on the linear combinations of the smallest compact localized states centered at different positions \cite{Rhim_Yang_2021}. 

\begin{figure}
    \centering
     a) \includegraphics[scale=0.65]{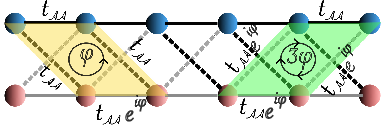}\\
     b) \includegraphics[scale=0.75]{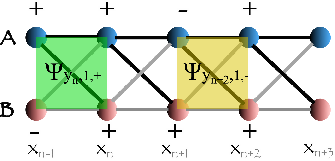}
    \caption{ 
    a) In this flat band regime, we observe that when two different closed paths are followed, the fermions acquire a phase difference of $\phi$ and $3 \phi$ with $\phi= \pi$, resulting in a $\pi-$ flux in the Kitaev-Creutz type ladder model in Eq. \ref{eq:Model.5}.
    b) The two CSLs in the flat band regime are represented by filled regions, with the signs of their component amplitudes indicated (see Eq. \ref{eq: FB.CLS.KSPACE.2}).
    }
    \label{fig:compact-localized-states-1}
\end{figure}

The eigenvalues of the Hamiltonian, as demonstrated by Eq. \ref{eq:Model.5}, have two topological flat bands (FB) with $\epsilon(k)= \pm 2$ when $\overline{\varepsilon}=0$ and $\lambda= \pm 1$, in this regime $\nu= \mp 1$ (see Fig. \ref{fig:phase-diagram}) . Furthermore, the electrons can obtain a phase difference of $\pi$ along closed trajectories (see Fig. \ref{fig:compact-localized-states-1} a)). The Bloch state creation operator for the FB is expressed as
\begin{eqnarray} \label{eq: FB.CLS.KSPACE.1}
\Psi_{k, \lambda, \pm}^{\dag}= \frac{1}{2} \left\{ \left( e^{i\lambda kl/2} \pm e^{-i \lambda kl/2} \right) a_{k}^{\dag}+ \left(e^{i\lambda kl/2} \mp e^{-i \lambda kl/2} \right) b_{k}^{\dag}\right\} 
 \end{eqnarray}
where the sign $\pm$ is for $\epsilon(k)= \pm 2$, respectively. 

The energy degeneracy means that any combination of the FB Bloch states is an eigenstate. Furthermore, the Fourier transform of these states is also an eigenstate. To illustrate this, let us calculate the Fourier transform of Eq. \ref{eq: FB.CLS.KSPACE.1}.

\begin{eqnarray} \label{eq: FB.CLS.KSPACE.2}
\fl \Psi^{\dag}_{\boldsymbol{y}_{n}, \lambda, \pm}= \mathcal{N} \int_{BZ} dk e^{i k  \boldsymbol{y}_{n}} \Psi_{k, \lambda, \pm}^{\dag}  = \frac{1}{2} \left(a_{n}^{\dag} \pm a_{n- \lambda}^{\dag}+ b_{n}^{\dag} \mp b_{n- \lambda}^{\dag} \right)
 \end{eqnarray}
where $y_{n}=x_{n}-l/2$ and $\mathcal{N}$ is a normalization constant. As shown in Fig. \ref{fig:compact-localized-states-1} b), $\Psi_{\boldsymbol{y}_{n}, \lambda, \pm}$ takes the form of a localized square plaquette centered $y_{n}$, that is, between cell $n$ and $n-1$.

One can verify that the sites of plaquettes obey the following relations, analogous to Landau levels states relations \cite{Liu_2014},

\begin{eqnarray} \label{eq: FB.CLS.KSPACE.3}
\fl        \sum_{\boldsymbol{y}_{n}} (-1)^{n} \Psi_{\boldsymbol{y}_{n},\lambda,+}^{(1)}+ \Psi_{\boldsymbol{y}_{n},\lambda,+}^{(2)}=0 \mbox{ and } \sum_{\boldsymbol{y}_{n}}  \Psi_{\boldsymbol{y}_{n},\lambda,-}^{(1)}+(-1)^{n} \Psi_{\boldsymbol{y}_{n},\lambda,-}^{(2)}=0
 \end{eqnarray}
with $\Psi^{(1,2)}$ as the first- and second- component of the FB Bloch state. A recent study \cite{Elias2023} has also demonstrated a connection between flat bands and Landau levels. Due to destructive interference, the electron is confined within the plaquette, resulting in a quenched kinetic energy that FB regulates.

\begin{figure}[t]
\fl a)\includegraphics[scale=0.6]{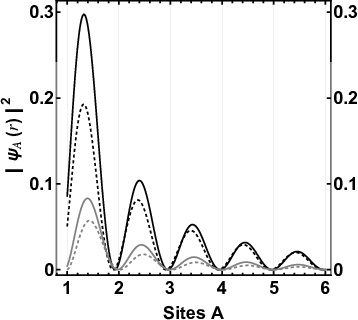}
b) \includegraphics[scale=0.6]{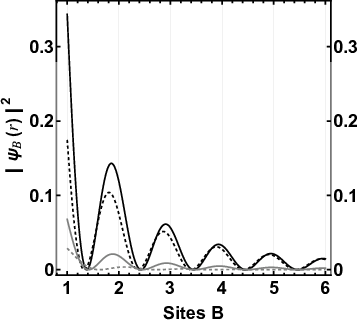}\\
\fl c) \includegraphics[scale=0.6]{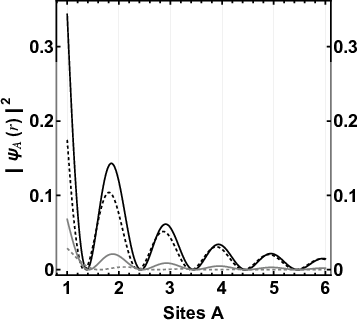}
d) \includegraphics[scale=0.6]{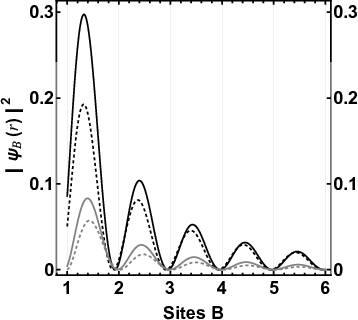}
   \caption{The electron density for sites $A$ and $B$ is shown in panels a),b) for the conduction band and in panels c), d) for the valence band. The used parameters were $\lambda=1$ and $\overline{\varepsilon}=0$ (solid black curve), $\overline{\varepsilon}=1$ (dashed black curve), $\overline{\varepsilon}=2$ (solid gray curve) and $\overline{\varepsilon}=3$ (gray dots). In analogy to Eq. \ref{eq: FB.CLS.KSPACE.2}, at sites $A$ and $B$ there is constructive and destructive interference in the conduction band case and vice versa in the valence band case. Notice how as  the system approaches the flat band case ($\overline{\varepsilon}=0$) the peaks tend to be more pronounced.
  }
    \label{fig:localized-eigenfunctions}
\end{figure}

 Let us now write a Bloch state as
\begin{equation} \label{eq:bloch_state_1-1}
    \ket{u_{k}^{\pm}}= \frac{1}{2} \left( e^{i \omega_{k}/2} \pm e^{- i \omega_{k}/2}, e^{i \omega_{k}/2} \mp e^{- i \omega_{k}/2}\right)^{T}.
\end{equation}
Therefore, the eigenstates in the real space are 
\begin{equation}
    \ket{\psi_{\pm}(x)}= \frac{l}{2 \pi} \int_{- \pi/l}^{\pi/l}  dk e^{ikx} \ket{u_{k}^{\pm}}\equiv (\psi_{A, \pm}(x),\psi_{B, \pm}(x))^{T}
\end{equation}

Fig. \ref{fig:localized-eigenfunctions} shows the eigenfunction in real space for sites $A$ and $B$ with different values of $\overline{\varepsilon}$, keeping $\lambda=1$. As can be seen in Fig. \ref{fig:localized-eigenfunctions} a) and b), there is constructive and destructive interference between sites $A$ and $B$, respectively. This is contrary to what is observed in Fig. \ref{fig:localized-eigenfunctions} c) and d). This is in accordance with Eq. \ref{eq: FB.CLS.KSPACE.2} due to the change of sign when considering a conduction and valence bands. Furthermore, when $\overline{\varepsilon}=0$, the electron density for the sites $A$ and $B$ is sharper than in any other of the scenarios; however, in this case, the spectrum is also highly degenerate so as explained in sec. \ref{sec:floppy_modes}, care must be taken in its interpretation.

\subsection{Equivalent SSH model\label{Sec: FlatBands}}

\begin{figure}[t]
    \centering
    \large{a)} \includegraphics[scale=0.7]{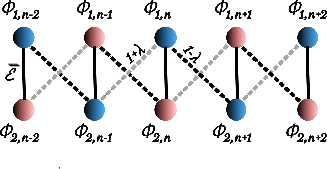}\\
    \large{b)}
    \includegraphics[scale=0.7]{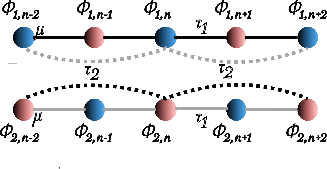}
    \caption{a) By transforming the basis (Eq. \ref{eq:17}) of the Kitaev-Creutz model, we obtain an equivalent ladder consisting of two SSH chains that are distinguished by red and blue sites (see the online version). b)This equivalent ladder can be divided into two chains with hopping of the first and second neighbors in Eq. \ref{eq:18}. Isolated atomic-like states, or CLS, are obtained at each site when $\lambda \to \pm 1$ and $\overline{\varepsilon}=0$ with $\tau_{1}=\tau_{2}=0$.}
    \label{fig:equivalent-models}
\end{figure}

We can observe a periodic chain with a two-sublattice structure in Fig. \ref{fig:sporbitals} b). Sites on sublattice $A$ have the diagonal hopping to the left, $t_{AB}$, and right $t_{BA}$, while for sites belonging to the other sublattice, it is just the opposite. According to the assumptions in Eq. \ref{eq:Model.3}, the time-independent Schrödinger equation can be transformed into equivalent difference equation forms for any pair of sites. Therefore, the following pairs of equations are valid for the given system.

\begin{eqnarray} \label{eq: 15}
 \fl (E- \varepsilon) \psi_{A, n}= t_{AA} \left( \psi_{A,n-1}+ \psi_{A,n+1} \right) + t_{AB} \left( \psi_{B,n+1}- \psi_{B,n-1} \right) \nonumber \\
\fl  (E+ \varepsilon) \psi_{B,n} = -t_{AA} \left( \psi_{B,n+1}+ \psi_{B,n-1} \right) + t_{AB} \left( \psi_{A,n-1}- \psi_{A,n+1} \right)
 \end{eqnarray}
To facilitate the analysis, we introduce the following basis transformation

\begin{equation} \label{eq:15.1}
\fl \left( \begin{array}{lcc}
\phi_{1,n}\\
\phi_{2,n}
\end{array} \right)= M \left( \begin{array}{lcc}
\psi_{A,n}\\
\psi_{B,n}
\end{array} \right) , \,\, M= \left( \begin{array}{lcc}
1 & 1\\
1 & -1
\end{array} \right)
\end{equation}

Then, from Eqs. \ref{eq: 15} and \ref{eq:15.1}, we obtain the following equations

\begin{eqnarray} \label{eq:17}
  \overline{E} \phi_{1,n} -\overline{\varepsilon} \phi_{2,n}=(1-\lambda) \phi_{2,n+1}+ (1+\lambda) \phi_{2,n-1} \nonumber  \\
    \overline{E} \phi_{2,n}- \overline{\varepsilon} \phi_{1,n}= (1+\lambda) \phi_{1,n+1}+ (1-\lambda) \phi_{1,n-1}
 \end{eqnarray}
where $\overline{E}= E/t_{AA}$. This equation system corresponds to two coupled SSH chains, as depicted in Fig. \ref{fig:equivalent-models} a).

\subsection{Supersymmetric transformation by squaring the Hamiltonian\label{sec:floppy_modes}}

In this subsection we will show that the square of the Hamiltonian decouples the sublattices  and renormalizes the hopping and on-site energies making it somewhat analogous to a phonon problem and akin to a supersymmetric transformation \cite{DikiMatsumoto2023,Yoshida2021,2023Mizoguchi,TomonariMizoguchi2022}.  To understand this, from Eq. \ref{eq:17}, we use the second equation in the first and vice versa. Then Eq. \ref{eq:17} can be rewritten as 
\begin{eqnarray} \label{eq:18}
\fl (\overline{E}^2- \mu) \phi_{j,n}= \tau_{1}(\phi_{j,n-1}+\phi_{j,n+1})  + \tau_{2} (\phi_{j,n+2}+\phi_{j,n-2}), \, j=1,2.
\end{eqnarray}
with $  \mu= \overline{\varepsilon}^{2}+2(1+ \lambda^{2})$, $\tau_{1}= 2 \overline{\varepsilon}$, and $ \tau_{2}= (1- \lambda^{2})$.

 Equation \ref{eq:18} is equivalent to square the Hamiltonian \ref{eq:Model.3}. It is also equivalent to remove one of the bipartite sublattices \cite{Naumis2007, Barrios-Vargas_2011,Leonardo2021}, in this case for the Kitaev-Creutz ladder leaving two  decoupled periodic chains with  nearest neighbour hoppings  ($\tau_{1}$), next-nearest hoppings ($\tau_{2}$) and with an effective on-site energy $\mu$ (See Fig. \ref{fig:equivalent-models} b)). The dispersion relations are obtained from Eq. \ref{eq:18} using a procedure similar to that exposed in Eq. \ref{eq: Results.A.4},
\begin{eqnarray} \label{eq: 20}
 \overline{E}^2= \mu+ 2 \tau_{1} \cos(kl)+ 2 \tau_{2} \cos(2 kl) \nonumber \\
      \Leftrightarrow  \overline{E}_{\pm}=\pm \sqrt{ (\overline{\varepsilon}+2 \cos(kl))^{2}+ 4 \lambda^{2} \sin^{2}(kl)} 
 \end{eqnarray}
or in a simpler form, by taking the square of the Hamiltonian \ref{eq:mostsimpleandfundamental} which reduces to 
\begin{equation} \label{eq:square_hamiltonian}
    \mathcal{H}^{2}(k)=  \left( \begin{array}{cc}
       \epsilon^{2}(k) & 0 \\
       0 &  \epsilon^{2}(k)
   \end{array} \right)
\end{equation}
The eigenvalues of $\mathcal{H}^{2}(k)$ are simply the square of those of $\mathcal{H}(k)$ explaining the hole-particle symmetry of the spectrum seen in Fig.  \ref{fig:energy-spectra}. We now observe that while any eigenfunction of $\mathcal{H}(k)$ is also an eigenfunction of $\mathcal{H}^{2}(k)$, the inverse in not necessarily true. Thus, the eigenfunctions of $\mathcal{H}(k)$ are sublattice polarized while those of $\mathcal{H}^{2}(k)$ are not necessarily polarized. 

In the flat band regime ($\overline{\varepsilon}=0, |\lambda|=1$), and from Eq. \ref{eq:18}, $\tau_{1}=\tau_{2}=0$ indicating the existence of localized atomic-like states that are equivalent to the CLS shown in Sec \ref{Sec: Analogous_Landa_Levels} (see Fig. \ref{fig:equivalent-models} b). Moreover, for the flat band 
$ \mathcal{H}^{2}(k)=(4)\ \mathbbm{1}_{2 \times 2}$ and the eigenfunctions are arbitrary linear combinations of the basis vectors $(1,0)^{T}$ and $(0,1)^{T}$.  This emphasizes the very peculiar localization properties of flat bands as seen in other systems \cite{Elias2023,Leonardo2022,Leonardo2023}. 
Therefore, the flat band now becomes a massive degenerate ground state of $\mathcal{H}^{2}(k)$.
As in other systems, the squared Hamiltonian can be interpreted as a massive vibrational band \cite{Leonardo2021} quite similar to the protected electronic boundary modes found in the QHE and topological insulators \cite{Kane2014} and which are well-known in the rigidity theory of glasses \cite{PHILLIPS1979153,Huerta2002,HuertaPRB_2002,Flores_2011}.


\subsection{ Topological properties II: \textit{Fubini-Study} metric}

The quantum geometry tensor is a key factor in understanding the behavior and characteristics of physical systems at the quantum level. It is particularly useful in the analysis of topological insulators and materials with flat bands. Moreover, it can be used to gain insight into the electronic structure and properties of these materials \cite{LEDWITH2021168646, TomokiOzawa2021Sep, TomokiOzawa2021Jul, TomokiOzawa2021Jul13}.

In general, we consider a quantum state $\ket{\psi(\boldsymbol{\xi})}$ in the $N$-dimensional parameter space, where $\boldsymbol{\xi}$ $=(\xi_{1}, \xi_{2}, \ldots, \xi_{N})$ is a set of parameters. Thus, this space can be endowed with the geometric quantum tensor \cite{TomokiOzawa2018, Aleksi2021, Cayssol_2021, Kruchkov2022, BernevigBogdan2022, TomokiOzawa2019}, given by

\begin{eqnarray} \label{eq: Quantum metric tensor}
    Q_{\mu \nu}(\boldsymbol{\xi}) \equiv \braket{\partial_{\mu}\psi(\boldsymbol{\xi})|\mathbb{P}_{\psi(\boldsymbol{\xi})}| \partial_{\nu} \psi(\boldsymbol{\xi})}
\end{eqnarray}
where $\mathbb{P}_{\psi(\boldsymbol{\xi})}$ is the orthogonal complement projector,
\begin{eqnarray}\label{eq: orthogonal complement projector}
    \mathbb{P}_{\psi(\boldsymbol{\xi})}= \mathbbm{1}-\ket{\psi(\boldsymbol{\xi})}\bra{\psi (\boldsymbol{\xi})}.
\end{eqnarray}
Since $Q_{\mu \nu}$ can have complex values, this leads to the \textit{FS} metric ($g_{\mu \nu}$), which is the real part of the quantum metric $g_{\mu \nu}(\boldsymbol{\xi})= \mbox{Re}[Q_{\mu \nu (\boldsymbol{\xi})}]$. The imaginary part is associated with the Berry curvature $\Omega_{\mu \nu}(\boldsymbol{\xi})$ and is given by 

\begin{eqnarray} \label{eq: relation between berry and imaginary part}
  \mbox{Im}[Q_{\mu \nu}]= - \frac{\Omega_{\mu \nu}}{2}.
\end{eqnarray}
The \textit{FS} metric measures the statistical distance between nearby pure quantum states $\ket{\psi(\boldsymbol{\xi})}$ and $\ket{\psi(\boldsymbol{\xi+ d \xi})}$, providing a means of distinguishing them \cite{bengtsson_zyczkowski_2006}. 

For calculating the \textit{FS} metric, it is necessary to consider the Bloch states, which are given by

\begin{eqnarray} \label{eq: bloch states for fubini}
 \ket{u^{+}_{k}}= (\cos(\omega_{k}/2), i \sin(\omega_{k}/2))^{T}, \mbox{ for conduction band} \nonumber \\
     \ket{u^{-}_{k}}= (i \sin(\omega_{k}/2),  \cos(\omega_{k}/2))^{T}, \mbox{ for valence band}.\\
 \end{eqnarray}
Then, the derivatives are
\begin{eqnarray} \label{eq: ApB.1}
\ket{\partial_{k}u^{+}_{k}}= \frac{\partial_{k} \omega_{k}}{2}(-\sin(\omega_{k}/2), i \cos(\omega_{k}/2))^{T},  \\
     \ket{\partial_{k}u^{-}_{k}}= \frac{\partial_{k} \omega_{k}}{2} (i \cos(\omega_{k}/2),  -\sin(\omega_{k}/2))^{T}. \\   
 \end{eqnarray}
and following the formula for the \textit{FS} metric, it can be proved that \cite{bengtsson_zyczkowski_2006,Cheng2010}
\begin{eqnarray} \label{eq: quantum metric for the model}
 g_{kk}^{\pm}= \braket{\partial_{k} u^{\pm}_{k}|\partial_{k}u_{k}^{\pm}}- \braket{\partial_{k} u_{k}^{\pm}|u_{k}^{\pm}}\braket{u_{k}^{\pm}| \partial_{k} u_{k}^{\pm}} = \left( \frac{\partial_{k} \omega_{k}}{2} \right)^{2}.
 \end{eqnarray}
The first observation is that from Eq. \ref{eq: quantum metric for the model} it follows
\begin{eqnarray} \label{eq: relation between fubini metric and winding density}
    \sqrt{\mbox{det}(g_{kk}^{\pm})}= \frac{|\partial_{k} \omega_{k}|}{2}.
\end{eqnarray}
We define the quantities $\mbox{vol(BZ)}$ and $\mbox{vol}(S^{1})$ given by

\begin{eqnarray} \label{eq: Brillouin Zone and S^1 volume}
\mbox{vol(BZ)} \equiv \int_{\mbox{BZ}} \sqrt{\mbox{det}(g_{kk}^{\pm})} dk, \,\, \mbox{Vol}(S^{1}) \equiv \int_{S^{1}} d \theta. 
 \end{eqnarray}
In the case where the \textit{FS} metric are nondegenerate, $\mbox{vol(BZ)}$ and $\mbox{vol}(S^{1})$ correspond to the \textit{volumes} of the Brillouin Zone (BZ) and the unitary circle $S^{1}$, respectively.
Thus, we obtained from Eqs. (\ref{eq:Results.B.1}, \ref{eq: relation between fubini metric and winding density}, \ref{eq: Brillouin Zone and S^1 volume}) the following inequalities
\begin{eqnarray} \label{eq: relation between winding number and brillouin zone}
 \pi |\nu| \leq \mbox{vol(BZ)} \leq \frac{1}{2} \mbox{vol}(S^{1}).
\end{eqnarray}
Therefore, the minimum value of the quantum volume is determined by the winding number of the occupied Bloch bundle. In the 2D case, this is equivalent to the result found by Ozawa and Mera, where the equalities are associated with a flat Kähler structure \cite{BrunoMeraZhang2022, TomokiOzawa2021Jul, TomokiOzawa2021Jul13, TomokiOzawa2021Sep}. This 2D version has also been used in topological superconductors to establish a relationship between the topological properties of the system and the superfluid weight, as demonstrated in \cite{Peotta2015,Peotta2016,Peotta2016dec,Bernevig2020,Jonah2022, JonahHerzog2022}.

\begin{figure}
    \centering
    \includegraphics[scale=0.4]{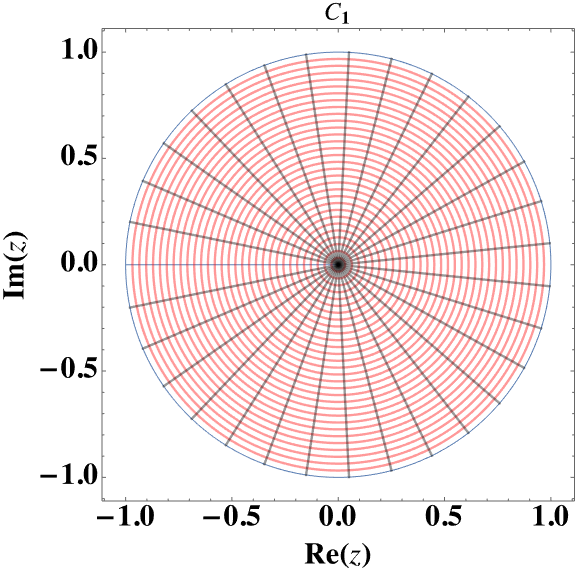}
    \caption{The complex plane $\mathbb{C}_{1}$  under the relation $\triangleright$ (Eq. \ref{eq: equivalence relation rays}) can be interpreted as each state $z$ on a specific external ray is an element of the equivalence class, $[z]$. This leads to an isomorphism between $\mathbb{C}_{1}$ and $S^{1}$ (blue circle), and the distance between two equivalence classes $[z_{1}]$ and $[z_{2}]$ is given by the angular difference between the respective external rays (cf. Eq. \ref{eq: arc length FS and S^1}). In this space, the distance between any two adjacent external rays depicted in the figure is the same. }
    \label{fig: S^1 under equivalence relation}
\end{figure}

\begin{figure}[t]
    \fl
   \large{a)} \includegraphics[scale=0.6]{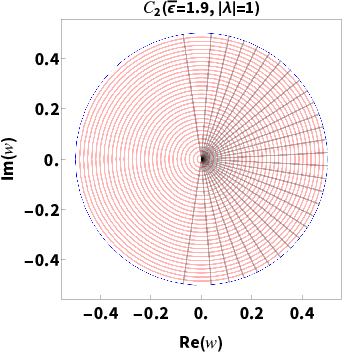}
   \large{b)} \includegraphics[scale=0.6]{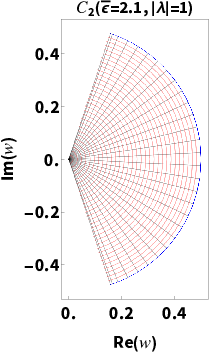}
    \caption{The image of $\mathbb{C}_{1}$ (Fig. \ref{fig: S^1 under equivalence relation}) under the mapping $f$ (Eq. \ref{eq: mapping f}) with the values $|\lambda|= 1 $ and a) $\overline{\varepsilon}=1.9$, b)$\overline{\varepsilon}=2.1$ are shown. In the nontrivial topological regime, $ |\overline{\varepsilon}| \leq 2$, $f$ maps rays and circles from $\mathbb{C}_{1}$ (Fig. \ref{fig: S^1 under equivalence relation}) to rays and circles with half the radius, respectively. On the other hand, in the trivial topological regime, circles are transformed into arcs of circles, providing another geometrical interpretation of Eq. \ref{eq:Results.B.1}. The inequality in Eq. \ref{eq: relation between winding number and brillouin zone} is visible in these figures. Thus, $\mbox{vol}(BZ)$ (cf. Eq. \ref{eq: Brillouin Zone and S^1 volume}) corresponds to the perimeter of the blue circle (or the arc circle), respectively; therefore, in the topological regime, the equality holds (a)), while in the nontopological regime (b)) the strict inequality holds.}
    \label{fig: FS in the trivial and nontrivial topological regime}
\end{figure}

\begin{figure}[t]
  \fl
   \large{a)} \includegraphics[scale=0.6]{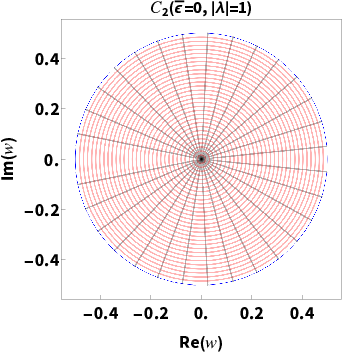} 
   \large{b)} \includegraphics[scale=0.6]{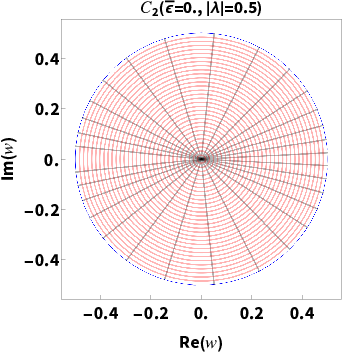}\\
 \fl  \large{c)} \includegraphics[scale=0.6]{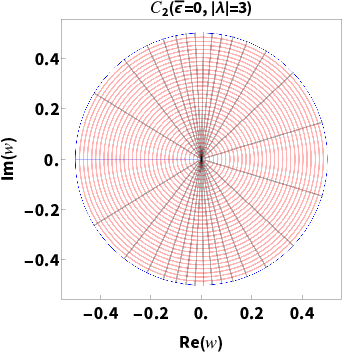} 
   \large{d)}   \includegraphics[scale=0.26]{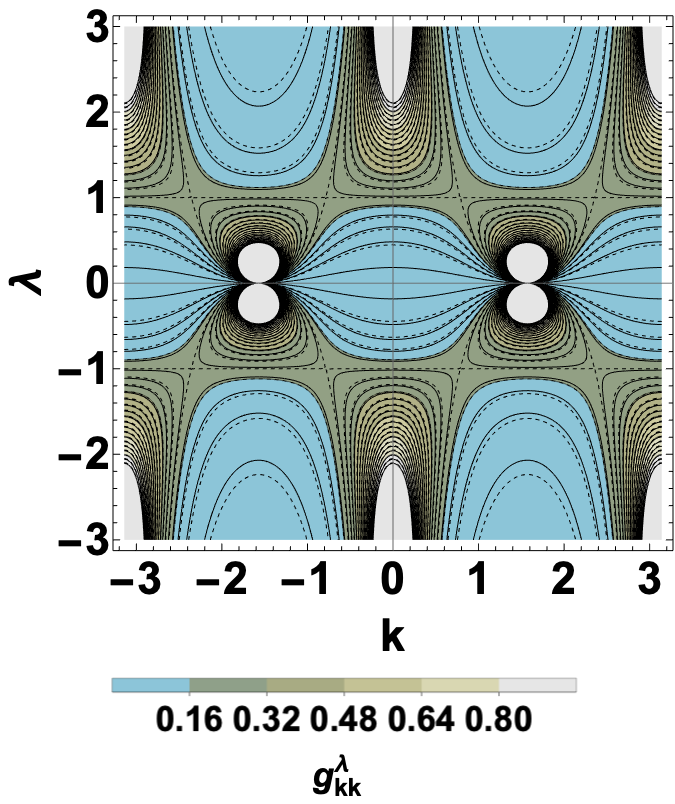}
    \caption{The image of $\mathbb{C}_{1}$ (Fig. \ref{fig: S^1 under equivalence relation}) under the mapping $f$ (Eq. \ref{eq: mapping f}) with the values $\overline{\varepsilon}=0$ and a) $|\lambda|= 1 $, b)$|\lambda|=0.5$ and c) $|\lambda|= 3 $, and a contour plot of the metric tensor $g_{kk}^{\pm}$ given by Eq. \ref{eq: quantum metric for the model} as a function of $\lambda$ and $k$ with $\overline{\varepsilon}=0$ (d) are shown. As discussed, $f$ preserves circles and external rays in the topological regime (see Fig. \ref{fig: FS in the trivial and nontrivial topological regime}); however, it deforms the distribution of rays, which is related to the FS metric tensor, $g_{kk}^{\pm}$. For example, in the case of c) where $|\lambda|= 3 $, the external rays appear to separate near the angles $k=0, \pm \pi$ and correspond to the maxima of $g_{kk}^{\pm}$ as seen in d). This geometric representation reveals that although all $3$ systems have a nontrivial topology with $ \nu =1$, the FS metric allows us to distinguish between them. In particular, system a), corresponding to the regime of flat bands, is the only system in which the distribution of rays is preserved under $f$.}
    \label{fig: FS space and their distance between states}
\end{figure}

On the other hand, to understand the geometry related to the \textit{FS} metric tensor, $g_{kk}^{\pm}$, it is necessary to consider the \textit{FS} arc element $ds_{FS}$ and the usual infinitesimal line elements in $S^1$, $ds_{S^{1}}$, which are given by 

\begin{eqnarray} \label{eq: arc length FS and S^1}
 ds_{FS}^{2}= g_{kk}^{\pm} dk^{2}= \left( \frac{1}{2} \right)^2 d \omega_{k}^2, \,\,\, |k| \leq \pi/l  \mbox{ and } ds^{2}_{S^{1}}= d \theta^{2}, \,\, |\theta| \leq \pi
 \end{eqnarray}
Thus, from Eq. \ref{eq: arc length FS and S^1}, we show that there is a locally conformal transformation between $S^1$ and the $FS$ manifold. Furthermore, we can conceive the space $FS$ and $S^1$ as embeddings in the complex plane with an additional structure given by the equivalence relation $\triangleright$ on $\mathbb{C}$, 
\begin{eqnarray} \label{eq: equivalence relation rays}
  z_{1} \triangleright z_{2} \mbox{ if } |z_{1}|=|R||z_{2}| \mbox{ and } \mbox{Arg}(z_{1})=\mbox{Arg}(z_{2})  
 \end{eqnarray}
where $|R|$ is a scale factor; i.e., $z_{1}$ and $z_{2}$ are on the same external ray. In addition,  let us consider the mapping $f$ defined by
\begin{eqnarray} \label{eq: mapping f}
 f: \,\mathbb{C}_{1} \to \mathbb{C}_{2}\nonumber \\
     z=r e^{i k} \mapsto w= f(z)= \frac{r}{2} e^{i \omega_{k}}
 \end{eqnarray}
 where the indexes refer to the first and second copies of the complex plane, respectively. Therefore, if we consider the usual infinitesimal line element on $\mathbb{C}$
 \begin{eqnarray} \label{eq: metric on C}
     ds^{2}_{\mathbb{C}} = dz dz^{*}= dr^2+ r^2 d{\theta}^{2},\,\, z=re^{i \theta} 
 \end{eqnarray}
 where ${}^{*}$ is the complex conjugate, we obtain that $\mathbb{C}_{1}$ and $\mathbb{C}_{2}$ under the relation $\triangleright$, are isomorphic to $S^1$ (see Fig. \ref{fig: S^1 under equivalence relation})  and $FS$ manifold (see Figs. \ref{fig: FS in the trivial and nontrivial topological regime}, \ref{fig: FS space and their distance between states}), respectively.

 Then, in Figs. \ref{fig: FS in the trivial and nontrivial topological regime} and \ref{fig: FS space and their distance between states} a)-c), we show the maps $\mathbb{C}_{1}$ onto $\mathbb{C}_{2}$ under $f$ given by Eq. \ref{eq: mapping f} with different values $\overline{\varepsilon}$ and $\lambda$; a first observation is that $f$ preserves circles and external rays only in the nontrivial topological regime, but deforms circles onto arcs of circles, given another geometrical perspective of Eq. \ref{eq:Results.B.1}. This mapping also allows for a visual realization of the inequality Eq. \ref{eq: relation between winding number and brillouin zone}, in which $\mbox{vol(BZ)}$ corresponds to the perimeter of the blue circle (or arc circle), respectively (see Fig. \ref{fig: FS in the trivial and nontrivial topological regime}).

 However, as we show in Fig. \ref{fig: FS space and their distance between states} a)-c), $f$ deform the distribution of rays, which is related to the FS metric tensor, $g_{kk}^{\pm}$. For example, in the case of Fig. \ref{fig: FS space and their distance between states} b) where $\lambda=0.5$, the external rays appear to separate near the angles $k= \pm \pi/2$ and correspond to the maxima of $g_{kk}^{\pm}$ as shown in Fig. \ref{fig: FS space and their distance between states} d). In a similar form,  in the case of Fig. \ref{fig: FS space and their distance between states} c) where $\lambda=|3|$ for $k=0, \pm \pi$. Something noteworthy about this geometric representation is that although all $3$ systems have a nontrivial topology with $|\nu|=1$, the FS metric allows us to establish differences between them. In particular, system Fig. \ref{fig: FS space and their distance between states} a) corresponding to the regime of flat bands is the only system in which the distribution of rays is not deformed under $f$.

 This can be confirmed as in the FB regime, the  \textit{FS} metric is constant as,
\begin{eqnarray} \label{eq: theta en FB}
    \partial_{k} \omega_{k}(\lambda= \pm 1)= \pm l \,\,\,  \mbox{(cf. \ref{eq:Results.A.3})}
\end{eqnarray}
and thus is equivalent to the usual metric over $S^{1}$. Such equivalence is readily found from the fact that an arc element $ds$ of a geodesic in the FS manifold is given by,
\begin{eqnarray}
    ds^2_{FS}=\left( \frac{l}{2} \right)^2 dk^{2} , \,\, |k| \leq \pi / l
\end{eqnarray}
equivalent to an arc element of a circle with a radius $1/2$.

\section{Conclusions} \label{Sec: Conclusions}
In conclusion, we have shown a mapping between a chain with $s-p$ orbitals onto a Kitaev- Creutz type model. With this map, we have found the existence of pseudo-Bogoliubov modes and compact localized states in the flat band (FB) regime, which obeys an analogous Landau relation (cf. Eq. \ref{eq: FB.CLS.KSPACE.3}). In addition, we obtained that there is a nontrivial topological transition by condition $\overline{\varepsilon}= \pm 2$ (see Fig. \ref{fig:phase-diagram}) that results in a nontrivial winding number. Furthermore, our analysis reveals the fingerprint of the flat bands with the \textit{FS metric} that allows us to distinguish between pure states of systems with the same topology; in particular, in this model, the \textit{ FS metric} is equivalent to the usual metric over $S^1$ in the FB regime. These findings could have implications for developing simpler models that aid understanding of flat band formation and its effects on many-body interactions.

\section*{Data availability statement}
All data that support the findings of this study are included within the article (and any supplementary files).

\section*{Acknowlegdgment}
A.E.C. thanks to the CONACyT scholarship for providing financial support. This work was supported by UNAM DGAPA PAPIIT IN102620 and CONACyT project 1564464.

\appendix
\section{Bogoulibov transformation \label{Sec:Appendix A}}
We define the pseudo-Bogoulibov modes $\gamma_{k}, \varrho_{k}$ as 

\begin{eqnarray} \label{eq:App.A.1}
 \gamma_{k}  \equiv u_{k} a_{k}+ v_{k} b_{k} \mbox{ and } \varrho_{k} \equiv -v_{k}^{*} a_{k}+ u_{k}^{*} b_{k}
 \end{eqnarray}
and the anti-commutation relations for $\gamma_{k}$ are
\begin{eqnarray} \label{eq: App.A.2}
\fl \{ \gamma_{k}, \gamma^{\dag}_{k'}\}= \left( u_{k} a_{k}+ v_{k} b_{k} \right) \left( u_{k'}^{*} a_{k'}^{\dag}+ v_{k'}^{*} b_{k'}^{\dag} \right) + \left( u_{k'}^{*} a_{k'}^{\dag}+ v_{k'}^{*} b_{k'}^{\dag} \right) \left( u_{k} a_{k}+ v_{k} b_{k} \right) \nonumber \\
\fl    = u_{k}u_{k'}^{*} \{a_{k}, a_{k'}^{\dag}\}+ v_{k} v_{k'}^{*} \{ b_{k}, b_{k'}^{\dag}\} + u_{k}v_{k'}^{*} \{a_{k}, b_{k'}^{\dag}\}+ u_{k'}^{*} v_{k} \{ b_{k}, a_{k'}^{\dag}\} \nonumber \\
\fl    = \left( u_{k} u_{k'}^{*}+ v_{k}v_{k'}^{*}\right) \delta_{kk'} \nonumber \\
\fl   = \delta_{kk'}, \mbox{ if } |u_{k}|^{2}+|v_{k}|^{2}=1
 \end{eqnarray}

\begin{eqnarray} \label{eq: App.A.3}
\fl \{ \gamma_{k}, \gamma_{k'}\}= \left( u_{k} a_{k}+ v_{k} b_{k} \right) \left( u_{k'} a_{k'}+ v_{k'} b_{k'} \right) + \left( u_{k'} a_{k'}+ v_{k'} b_{k'} \right) \left( u_{k} a_{k}+ v_{k} b_{k} \right) \nonumber \\
\fl   = u_{k}u_{k'} \{a_{k}, a_{k'}\}+ v_{k} v_{k'} \{ b_{k}, b_{k'}\} + u_{k}v_{k'} \{a_{k}, b_{k'}\}+ u_{k'} v_{k} \{ b_{k}, a_{k'}\} \\
 \fl   =0  
 \end{eqnarray}

\begin{eqnarray} \label{eq: App.A.4}
\fl  \{ \gamma_{k}^{\dag}, \gamma_{k'}^{\dag}\}= \left( u_{k}^{*} a_{k}^{\dag}+ v_{k}^{*} b_{k}^{\dag} \right) \left( u_{k'}^{*} a_{k'}^{\dag}+ v_{k'}^{*} b_{k'}^{\dag} \right)  + \left( u_{k'}^{*} a_{k'}^{\dag}+ v_{k'}^{*} b_{k'}^{\dag} \right) \left( u_{k}^{*} a_{k}^{\dag}+ v_{k}^{*} b_{k}^{\dag} \right)\nonumber \\
\fl    = u_{k}^{*}u_{k'}^{*} \{a_{k}^{\dag}, a_{k'}^{\dag}\}+ v_{k}^{*} v_{k'}^{*} \{ b_{k}^{\dag}, b_{k'}^{\dag}\}  + u_{k}^{*}v_{k'}^{*} \{a_{k}^{\dag}, b_{k'}^{\dag}\}+ u_{k'}^{*} v_{k}^{*} \{ b_{k}^{\dag}, a_{k'}^{\dag}\}\\
 \fl   =0
 \end{eqnarray}
and similar anticommutation relations for $\varrho_{k}$ under the interchange of $u_{k} \leftrightarrow -v_{k}^{*}$ and $v_{k} \leftrightarrow u_{k}^{*}$.
From the definition, \ref{eq:App.A.1} we can obtain the following result
\begin{eqnarray} \label{eq:App.A.5}
\fl \epsilon(k) \left( \gamma_{k}^{\dag} \gamma_{k}- \varrho_{k}^{\dag} \varrho_{k} \right)= \epsilon(k) \left[ (u_{k}^{*} a_{k}^{\dag}+ v_{k}^{*} b_{k}^{\dag})(u_{k} a_{k}+ v_{k} b_{k}) -(-v_{k} a_{k}^{\dag}+ u_{k} b_{k}^{\dag})(-v_{k}^{*} a_{k}+ u_{k}^{*} b_{k}) \right]  \nonumber\\
\fl    = \epsilon(k) \left[ \left(|u_{k}|^{2}-|v_{k}|^{2}  \right)a_{k}^{\dag} a_{k}+ 
     -\left(|u_{k}|^{2}-|v_{k}|^{2}  \right)b_{k}^{\dag} b_{k}+2u_{k}^{*} v_{k} a_{k}^{\dag} b_{k} +2 u_{k} v_{k}^{*} b_{k}^{\dag} a_{k} \right]
 \end{eqnarray}
If we choose $u_{k}$ and $v{k}$ as $ u_{k} \equiv \cos \left(\frac{\omega_{k}}{2} \right), v_{k} \equiv - i \sin\left(\frac{\omega_{k}}{2}\right)$ where
\begin{eqnarray} \label{eq: App.A.6}
 \sin(\omega_{k}) \equiv \frac{2 \lambda \sin(k l)}{ \epsilon(k)},  \,\,  \cos(\omega_{k}) \equiv \frac{\overline{\varepsilon}+2\cos(k l)}{ \epsilon(k)},\nonumber \\
 \epsilon(k)= \sqrt{(\overline{\varepsilon}+2 \cos(kl))^{2}+(2 \lambda \sin(kl))^{2}}.
 \end{eqnarray}
We recover the Hamiltonian form of Eq. \ref{eq:Model.5}. We note that only in the two cases $\overline{\varepsilon}=0, \lambda = \pm 1$, the relation $\omega_{k}$ with $k$ is linear, that is, $\omega_{k}= \pm kl, \mbox{ if } \lambda= \pm 1 \mbox{ and } \overline{\varepsilon}=0.$

\newpage
\section*{References}
\bibliographystyle{iopart-num}
\providecommand{\newblock}{}

\end{document}